\documentclass[aps,twocolumn,pra,preprintnumbers,showpacs,tightenlines]{revtex4}
\usepackage{amssymb}
\usepackage{epsfig,graphicx,times}

\input{tcilatex}
\begin{document}

\title{Measurement-based quantum phase estimation algorithm for finding
eigenvalues of non-unitary matrices}
\author{Hefeng Wang$^{1, 3}$, Lian-Ao Wu$^{2}$, Yu-xi Liu$^{1, 4}$, and
Franco Nori$^{1, 3}$}
\affiliation{$^{1}$Advanced Science Institute, The Institute of Physical and Chemical
Research~(RIKEN), Wako-shi, Saitama 351-0198, Japan\\
$^{2}$Department of Theoretical Physics and History of Science, The Basque
Country University~(EHU/UPV), Post Office Box 644, ES-48080 Bilbao, Spain
and IKERBASQUE, Basque Foundation for Science, ES-48011 Bilbao, Spain\\
$^{3}$Department of Physics, The University of Michigan, Ann Arbor, Michigan
48109-1040, USA\\
$^{4}$Institute of Microelectronics and Tsinghua National Laboratory for
Information Science and Technology~(TNList), Tsinghua University, Beijing
100084, China}

\begin{abstract}
We propose a quantum algorithm for finding eigenvalues of non-unitary
matrices. We show how to construct, through interactions in a quantum system
and projective measurements, a non-Hermitian or non-unitary matrix and
obtain its eigenvalues and eigenvectors. This proposal combines ideas of
frequent measurement, measured quantum Fourier transform, and quantum state
tomography. It provides a generalization of the conventional phase
estimation algorithm, which is limited to Hermitian or unitary matrices.
\noindent
\end{abstract}

\maketitle

\section{Introduction}

One of the most important tasks for a quantum computer would be to
efficiently obtain eigenvalues and eigenvectors of high-dimensional
matrices. It has been suggested~\cite{abrams} that the quantum phase
estimation algorithm~(PEA)~\cite{Kitaev95} can be used to obtain eigenvalues
of a Hermitian matrix or Hamiltonian. For a quantum system with a
Hamiltonian $H$, a phase factor, which encodes the information of
eigenvalues of $H$, is generated via unitary evolution $U=\exp (-iH\tau )$.
By evaluating the phase, we can obtain the eigenvalues of $H$. The
conventional PEA consists of four steps: preparing an initial approximated
eigenstate of the Hamiltonian $H$, implementing unitary evolution operation,
performing the inverse quantum Fourier transform~(QFT), and measuring binary
digits of the index qubits.

The PEA is at the heart of a variety of quantum algorithms, including Shor's
factoring algorithm~\cite{chuang}. A number of applications of PEA have been
developed, including generating eigenstates associated with an operator~\cite%
{tm}, evaluating eigenvalues of differential operators~\cite{szk}, and it
has been generalized using adaptive measurement theory to achieve a
quantum-enhanced measurement precision at the Heisenberg limit~\cite{hig}.
The PEA with delays considering the effects of dynamical phases has also
been discussed~\cite{wei}. The implementation of an iterative quantum phase
estimation algorithm with a single ancillary qubit is suggested as a
benchmark for multi-qubit implementations~\cite{mir}. The PEA has also been
applied in quantum chemistry to obtain eigenenergies of molecular systems~%
\cite{aa, whf}. This application has been demonstrated in a recent
experiment~\cite{expt}. Moreover, several proposals have been made to
estimate the phase of a quantum circuit~\cite{Ngpra}, and the use of phase
estimations for various algorithms~\cite{Ng1, Ng2}, including factoring and
searching.

The conventional PEA is only designed for finding eigenvalues of either a
Hermitian or a unitary matrix. In this paper, we propose a measurement-based
phase estimation algorithm~(MPEA) to evaluate eigenvalues of \emph{non}%
-Hermitian matrices. This provides a potentially useful generalization of
the conventional PEA. Our proposal uses ideas from conventional PEA,
frequent measurement, and techniques in one-qubit state tomography. This
proposal can be used to design quantum algorithms apart from those based on
the standard unitary circuit model. The proposed quantum algorithm is
designed for systems with large dimension, when the corresponding classical
algorithms for obtaining the eigenvalues of the non-unitary matrices become
so expensive that it is impossible to implement on a classical computer.

The structure of this work is as follows: In Sec.~\ref{const} we introduce
how to construct a non-Hermitian evolution matrix for a quantum system. In
Sec.~\ref{mpea}, we present the measurement-based phase estimation
algorithm, introducing two methods for obtaining the complex eigenvalues of
the non-Hermitian evolution matrix. We give two examples for the application
of MPEA and discuss how to construct Hamiltonian for performing the
controlled unitary operation in Sec.~\ref{example}. In Sec.~\ref{diss}, we
discuss the success probability of the algorithm and the efficiency of
constructing the non-Hermitian matrix. We close with a conclusion section.

\section{Constructing non-unitary matrices}

\label{const}

Now we describe how to construct non-unitary matrices on a quantum system. A
bipartite system, composed of subsystems \textit{A} and \textit{B}, evolves
under Hamiltonian 
\begin{equation}
H=H_{A}+H_{B}+H_{AB},
\end{equation}%
where $H_{A~(B)}$ is the Hamiltonian of subsystem \textit{A}(\textit{B}) and 
$H_{AB}$ is their interaction. We prepare the initial state of subsystem 
\textit{A} in its pure state $|\varphi _{A}\rangle \langle \varphi _{A}|$
and the initial state of subsystem \textit{B} in an arbitrary state $\rho
_{B}$. Then at time $t=0$, the state of the system is $\rho _{0}=|\varphi
_{A}\rangle \langle \varphi _{A}|\otimes \rho _{B}$. Let the system evolve
under the Hamiltonian $H$ for a time interval $\tau $, if subsystem \textit{A%
} is subject to a projective measurement $M=|\varphi _{A}\rangle \langle
\varphi _{A}|$ applied at time interval $\tau $, this is equivalent to
driving subsystem \textit{B} with an evolution matrix 
\begin{equation}
V_{B}(\tau )\equiv \langle \varphi _{A}|\exp \left( -iH\tau \right) |\varphi
_{A}\rangle .
\end{equation}%
This evolution matrix is in general neither unitary nor Hermitian.

The Hamiltonian $H$ of the whole quantum system can be spanned as: 
\begin{equation}
H=\sum_{j=1}^{D}E_{j}|E_{j}\rangle \langle E_{j}|,
\end{equation}%
with eigenenergies $E_{j}$, and the corresponding eigenvectors $%
|E_{j}\rangle $ can be spanned in terms of tensor products of basis vectors
\{$|\psi _{k}^{A}\rangle $\} and \{$|\psi _{r}^{B}\rangle $\} of Hilbert
spaces of subsystems \textit{A} and \textit{B}, which are of dimensions $%
D_{A}$ and $D_{B}$, respectively, and $D=D_{A}\times D_{B}$. Using the bases
for \textit{A} and \textit{B}, we have 
\begin{equation}
|E_{j}\rangle =\sum_{k=1}^{D_{A}}\sum_{r=1}^{D_{B}}f_{kr}^{j}|\psi
_{k}^{A}\rangle \otimes |\psi _{r}^{B}\rangle ,
\end{equation}%
and the evolution matrix on subsystem \textit{B}, after the measurement $M$
performed on subsystem \textit{A }at time interval $\tau $, becomes 
\begin{equation}
V_{B}(\tau )=\langle \varphi ^{A}|e^{-iH\tau }|\varphi ^{A}\rangle
=\sum_{r,s=1}^{D_{B}}V_{rs}|\psi _{r}^{B}\rangle \langle \psi _{s}^{B}|,
\end{equation}%
where%
\begin{equation}
V_{rs}=\sum_{j=1}^{D}e^{-iE_{j}\tau
}\sum_{k,l=1}^{D_{A}}f_{kr}^{j}f_{ls}^{j\ast }c_{k}^{A}c_{l}^{A\ast },
\end{equation}%
and 
\begin{equation}
c_{k}^{A}=\langle \varphi ^{A}|\psi _{k}^{A}\rangle .
\end{equation}

More generally, we can construct different evolution matrices by performing
measurements on subsystem \textit{A} with different time intervals and/or
different measurement bases. For example, by sequentially performing
projective measurements with time intervals $\tau _{1}$, $\tau _{2}$, $\tau
_{3}$, an evolution matrix 
\begin{equation}
V_{B}(\tau _{1},\tau _{2},\tau _{3})=V_{B}(\tau _{3})V_{B}(\tau
_{2})V_{B}(\tau _{1})
\end{equation}%
is constructed. We can also combine unitary evolution matrices with the
non-unitary transformations on subsystem \textit{B} to construct some
desired evolution matrices.

\section{Measurement-based quantum phase estimation algorithm}

\label{mpea}

For the bipartite system, set the initial state of the system in a separable
state 
\begin{equation}
\rho _{0}=|\varphi _{A}\rangle \langle \varphi _{A}|\otimes \rho _{B},
\end{equation}%
and let the system evolve under the Hamiltonian in Eq.~($1$). Then after
performing $m$ successful projective measurements on subsystem \textit{A}
with time intervals $\tau $, the evolution on the Hilbert space of subsystem 
\textit{B} is driven by $\left[ V_{B}(\tau )\right] ^{m}$, and the state of
subsystem \textit{B} evolves to~\cite{nakazato}%
\begin{equation}
\rho _{B}^{(\tau )}(m)=\frac{\bigl[V_{B}(\tau )\bigr]^{m}\rho _{B}\bigl[%
V_{B}^{\dagger }(\tau )\bigr]^{m}}{P^{\left( \tau \right) }\left( m\right) },
\end{equation}%
where 
\begin{equation}
P^{\left( \tau \right) }\left( m\right) =\text{Tr}_{B}\bigl\{\lbrack
V_{B}(\tau )]^{m}\rho _{B}[V_{B}^{\dagger }(\tau )]^{m}\bigr\}
\end{equation}%
is the probability of finding subsystem \textit{A} still in state $|\varphi
_{A}\rangle $ after each of the $m$ measurements.

The evolution matrix $V_{B}(\tau )$ can be spanned as 
\begin{equation}
V_{B}(\tau )=\sum_{k}\lambda _{k}|u_{k}\rangle \langle v_{k}|,
\end{equation}%
where $|u_{k}\rangle $ and $\langle v_{k}|$ are the right- and
left-eigenvectors of $V_{B}(\tau )$\ and $\lambda _{k}$ is the corresponding
eigenvalue~\cite{nakazato} satisfying $0\leq |\lambda _{k}|\leq 1$. In the
large $m$ limit, the operator $[V_{B}(\tau )]^{m}$\ is dominated by a single
term $\lambda _{\max }^{m}|u_{\max }\rangle \langle v_{\max }|$, provided
the largest eigenvalue $\lambda _{\max }$\ is unique, discrete, and
non-degenerate. In the limit of large $m$ and finite $\tau $, $\rho
_{B}^{(\tau )}(m)$ tends to a pure state, independent of the initial~(mixed)
state of subsystem \textit{B}. The final state of $\rho _{B}^{(\tau )}(m)$\
is dominated by $|u_{\max }\rangle $, and this outcome is found with
probability~\cite{nakazato} 
\begin{equation}
P^{\left( \tau \right) }\left( m\right) \longrightarrow |\lambda _{\max
}|^{2m}\langle u_{\max }|u_{\max }\rangle \langle v_{\max }|\rho
_{B}|v_{\max }\rangle .
\end{equation}%
The state of subsystem \textit{B} evolves to $|u_{\max }\rangle $, after
performing a number of operations of $V_{B}(\tau )$. Then we can evaluate $%
\lambda _{\max }$ by resolving the phase of the state. If we prepare the
initial state of subsystem \textit{B} in a pure initial state that is close
to an eigenstate of the matrix $V_{B}(\tau )$, the state of the subsystem 
\textit{B} can evolve to other eigenstates of $V_{B}$. Then we can also
obtain the corresponding eigenvalues of $V_{B}(\tau )$.

Based on the above analysis, we suggest a measurement-based phase estimation
algorithm for evaluation of the eigenvalues of the matrix $V_{B}(\tau )$. As
in the circuit shown in Fig.~$1$($a$), three quantum registers are prepared.
From top to bottom: an index register, a target register and an interacting
register. The index register is a single qubit, which is used as control
qubit and to readout the final results for the eigenvalues; the target
register is used to represent the state $\rho _{B}$ of subsystem \textit{B};
and the interacting register represents the state $|\varphi _{A}\rangle $\
of subsystem \textit{A}.

The initial state of the circuit is prepared in the state%
\begin{equation}
|0\rangle \langle 0|\otimes \rho _{B}\otimes |\varphi _{A}\rangle \langle
\varphi _{A}|,
\end{equation}%
with subsystem \textit{A} in a pure state $|\varphi _{A}\rangle \langle
\varphi _{A}|$, and subsystem \textit{B} in state $\rho _{B}$. The
construction of the controlled evolution matrix $V_{B}(\tau )$ on the target
register is achieved by implementing the controlled unitary~($C$-$U$)
transformation for the whole quantum system and successfully performing the
projective measurement $M=|\varphi _{A}\rangle \langle \varphi _{A}|$ on the
interacting register with time interval $\tau $. Note here for the unitary
transformation $U=\exp \left( {-iHt}\right) $, we set $t$ such that $m$
projective measurements are performed successfully on subsystem $A$ at\ the
time interval $\tau $, while the unitary transformation of the whole system
evolves for time period $t$. After performing $m$ successful periodic
measurements on the interacting register with time intervals $\tau $, as
shown in Fig.~$1$($b$), the state of the system is transformed to 
\begin{eqnarray}
&&\frac{1}{2}\biggl\{|0\rangle \langle 0|\otimes \rho_{B}+|1\rangle \langle
1|\otimes \bigl[V_{B}(\tau )\bigr]^{m}\rho _{B}\bigl[V_{B}^{\dagger }(\tau )%
\bigr]^{m}\biggr\}  \nonumber \\
&&\otimes |\varphi _{A}\rangle \langle \varphi _{A}|.
\end{eqnarray}%
The dominant term of this is 
\begin{equation}
\frac{1}{\sqrt{2}}\biggl[|0\rangle +(\lambda _{\max })^{m}|1\rangle \biggr]%
|u_{\max }\rangle |\varphi _{A}\rangle ,
\end{equation}%
and the state of the index qubit is dominated by 
\begin{equation}
|\psi _{\text{ind}}\rangle =\frac{1}{\sqrt{2}}\biggl[|0\rangle +(\lambda
_{\max })^{m}|1\rangle \biggr].
\end{equation}%
In general, $\lambda _{\max }$ is a complex number and can be written as 
\begin{equation}
\lambda _{\max }=\exp \left( i\varphi \right) =\exp \left[ i\left(
a+ib\right) \right] .
\end{equation}%
We can obtain $\lambda _{\max }$\ by resolving the phase factor 
\begin{equation}
\varphi =(a+ib).
\end{equation}

Two approaches can be used to resolve $\lambda _{\max }$: $\left( i\right) $
using single-qubit quantum state tomography~(QST)~\cite{kwiat}, and $\left(
ii\right) $\ using the measured quantum Fourier transform~(mQFT) combined
with projective measurements on a single qubit. The details of these two
approaches are given below.

\subsection{Approach using single-qubit state tomography}

Quantum state tomography can fully characterize the quantum state of a
particle or particles through a series of measurements in different bases~%
\cite{kwiat, liu}. In the approach using QST to resolve the eigenvalue of
the matrix $V_{B}(\tau )$, we prepare a large number of identical copies of
the state on the index qubit $|\psi _{\text{ind}}\rangle $, as shown in Eq.~(%
$17$), by running the MPEA circuit a number of times. Then the value of $%
\lambda _{\max }$ can be obtained by determining the index qubit state.

The state of the index qubit in Eq.~($17$) can be written as 
\begin{equation}
|\psi _{\text{ind}}\rangle =\frac{1}{\sqrt{2}}\bigl( |0\rangle +\exp \left[
m\left( -b+ia\right) \right] |1\rangle \bigr) .
\end{equation}%
In the QST approach, we perform a projective measurement on the index qubit
in the basis $|1\rangle \langle 1|$ to obtain the probability of finding the
index qubit in state $|1\rangle $, thus obtaining the value of $b$. With the
knowledge of $b$, then perform a $\pi /2$ rotation around the $x$-axis and a
measurement in the basis of the Pauli matrix $\sigma _{z}$ on the index
qubit, we can obtain the observable 
\begin{equation}
\langle \psi _{\text{ind}}|\exp \left( -i\frac{\pi }{4}\sigma _{x}\right)
\sigma _{z}\exp \left( i\frac{\pi }{4}\sigma _{x}\right) |\psi _{\text{ind}%
}\rangle
\end{equation}%
and thus obtain the value of $a$.

The measurement errors of QST, from counting statistics, obey the central
limit theorem. To obtain more accurate results, we have to prepare a larger
ensemble of the single qubit states.

\begin{figure}[tbp]
\includegraphics[width=\columnwidth, clip]{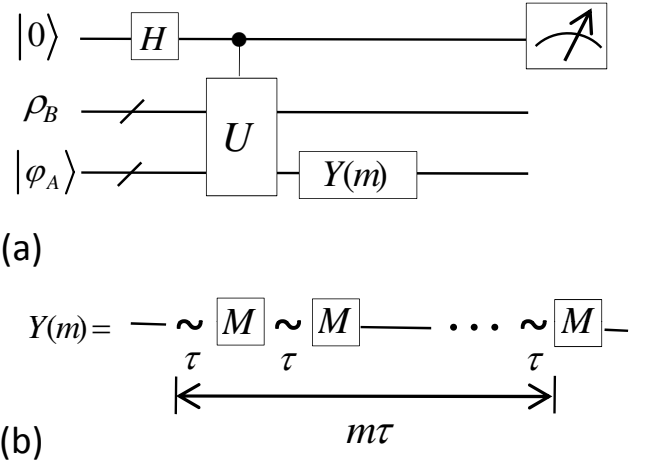}
\caption{Quantum circuit for measurement-based phase estimation
algorithm~(MPEA) using quantum state tomograph~(QST) approach. Part~($a$)
shows the circuit for the MPEA using QST. From top to bottom, an index
register, a target register and an interacting register are prepared in the
states $|0\rangle$, $\protect\rho_{B}$, and $|\protect\varphi_{A}\rangle$,
respectively. The index register is a single qubit used as control qubit;
the target register is used to represent the state of subsystem \textit{B};
and the interacting register is used to represent the state of subsystem 
\textit{A,} which interacts with \textit{B}. Part~($b$) shows the circuit
for performing $m$ projective measurements with period $\protect\tau $. In
the circuit, the unitary transformation $U=\exp(-iHt)$, we set $t$ such that 
$m $ projective measurements are performed successfully on subsystem \textit{%
A} separated by a time interval $\protect\tau $, while the unitary
transformation of the whole system evolves for time $t$.}
\end{figure}

\subsection{Approach using measured quantum Fourier transform combined with
projective measurements}

In the second approach, we use the techniques of measured quantum Fourier
transform and projective measurements to resolve the eigenvalue of the
matrix $V_{B}(\tau )$. The phases which encode the eigenvalues of $%
V_{B}(\tau )$ are in general complex numbers; the inverse QFT can be used to
resolve the real part $a$ of the phase $\varphi =(a+ib)$. The imaginary part 
$b$\ of the phase factor, $\varphi $, can be obtained by performing
single-qubit projective measurements. The details of this method are
discussed below.

In order to resolve $a$ up to $n$ binary digits using the inverse QFT, one
has to construct a series of controlled evolution matrices, $C$-$V_{B}(\tau
) $, in successive binary powers, from $(n-1)$ to $0$. In the MPEA, this is
done by implementing the $C$-$U$ operation on the whole system and
performing a series of $2^{k}$\ periodic measurements separated by time
intervals $\tau $ for $k=(n-1),(n-2),\cdots ,0$, respectively. The $C$-$U$
operation\ evolves for a time $t$, during which all the measurements are
performed successfully on the interacting register. Then we can obtain a
series of controlled transformation matrices in binary powers, $C$-$%
[V_{B}(\tau )]^{2^{k}},$ $k=(n-1),$ $(n-2),\cdots ,$ $0$. In Fig.~$2$($c$),
we show the circuit for the $k$th projective measurement with period $\tau $%
, $W(k)$, where the measurement $M=|\varphi _{A}\rangle \langle \varphi
_{A}| $ is performed $2^{k}$ times with period $\tau $ on the interacting
register, while the whole system evolves under the controlled unitary
operation $U$. The measurements on the interacting register are sequentially
performed $W(n-1),W(n-2),\cdots ,W(0)$. Then, correspondingly, on the index
qubit, we obtain single qubit states as 
\begin{eqnarray}
&&\frac{1}{\sqrt{2}}\biggl[|0\rangle +(\lambda _{\max
})^{2^{(n-1)}}|1\rangle \biggr],\frac{1}{\sqrt{2}}\biggl[|0\rangle +(\lambda
_{\max })^{2^{(n-2)}}|1\rangle \biggr],  \nonumber \\
&&\cdots ,\frac{1}{\sqrt{2}}\bigl(|0\rangle +\lambda _{\max }|1\rangle \bigr)%
.
\end{eqnarray}%
Afterward, we can retrieve $n$ binary digits of the real part $a$ of the
phase factor $\varphi $, of $\lambda _{\max }$, by performing a mQFT.

The mQFT technique implements a QFT using only a single qubit~\cite{PP, niu}%
. It uses the fact that gates within the Fourier transform are applied
sequentially on qubits. This modification of the QFT algorithm preserves the
probabilities of all measurements~\cite{niu}. The procedure for obtaining
the real part of the phase factor of $\lambda _{\max }$ by using mQFT is
shown in Fig.~$2$($a$), where the circuit in the dotted square is replaced
by circuits in the dotted squares shown in Fig.~$2$($b$), sequentially
obtaining $n$ binary digits of $a$. The details of this procedure are shown
below.

The initial state of the MPEA circuit is prepared as in Eq.~($14$). After
performing the $k$th periodic measurements $W(k)$ on the interacting
register, the dominant term of the state of the system becomes 
\begin{equation}
\frac{1}{\sqrt{2}}\bigg[|0\rangle +\lambda _{\max }^{2^{k}}|1\rangle \bigg]%
|u_{\max }\rangle |\varphi _{A}\rangle .
\end{equation}%
The state of the index qubit can be written as 
\begin{eqnarray}
|\psi _{\text{ind}}\rangle &=&\frac{1}{\sqrt{2}}\biggl[|0\rangle +\exp
\left( i(a+ib)2^{k}\right) |1\rangle \biggr]  \nonumber \\
&=&\frac{1}{\sqrt{2}}\biggl[|0\rangle +\exp \left( -b2^{k}\right) \exp
\left( ia2^{k}\right) |1\rangle \biggr].
\end{eqnarray}%
In order to resolve the real part $a$ of the phase factor, $\varphi $, we
first need to obtain the value of $b$, the imaginary part of the phase
factor. This can be achieved by using a single-qubit projective measurement.
One can prepare an ensemble of $|\psi _{\text{ind}}\rangle $ and perform
projective measurements $|1\rangle \langle 1|$. The value of $b$ can be
obtained through the probability for observing the index qubit in state $%
|1\rangle $.

Let 
\begin{equation}
r_{k}=\frac{1}{\sqrt{2}}\biggl[1+\exp \left( -b2^{k+1}\right) \biggr]^{1/2},
\end{equation}%
and let us run MPEA again and perform a single qubit operation $Q_{k}$ on
the index qubit such that the index qubit is rotated to state 
\begin{equation}
|\psi _{\text{ind}}^{\prime }\rangle =r_{k}\frac{1}{\sqrt{2}}\biggl[%
|0\rangle +\exp \left( ia2^{k}\right) |1\rangle \biggr],
\end{equation}%
where the single-qubit operation $Q_{k}$ is defined as 
\begin{equation}
Q_{k}=q_{k}\left( 
\begin{array}{cc}
1+e^{b2^{k}} & e^{-ia2^{k}}\left( 1-e^{b2^{k}}\right) \\ 
e^{ia2^{k}}\left( e^{b2^{k}}-1\right) & 1+e^{b2^{k}}%
\end{array}%
\right) ,
\end{equation}%
where $q_{k}=1/\sqrt{2\left[ 1+\exp \left( b2^{k+1}\right) \right] }$. Then
we apply the mQFT technique to resolve the real part $a$ of the phase factor 
$\varphi =(a+ib)$. We therefore obtain the eigenvalue $\exp \bigl[i(a+ib)%
\bigr]$ of the matrix $V_{B}(\tau )$.

In the MPEA, the $n$th binary digit of the phase factor is retrieved first,
and the partial measurement on the interacting register is performed in
sequence of $2^{n-1},2^{n-2}$ to $2^{0}$ times. This procedure provides high
fidelity for the state of the target register since each measurement drives
the state of the target register closer to $|u_{\max }\rangle $, the
eigenstate of $V_{B}(\tau )$. 
\begin{figure}[tbp]
\includegraphics[width=\columnwidth, clip]{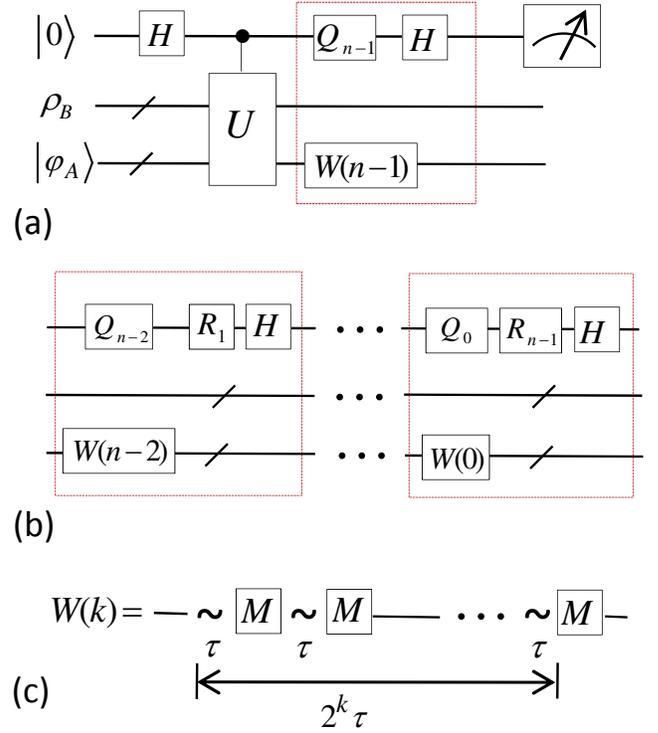}
\caption{Quantum circuit for a measurement-based phase estimation
algorithm~(MPEA) using measured quantum Fourier transformation. Part~(a)
shows the circuit for the MPEA. From top to bottom, an index register, a
target register and an interacting register are prepared in the states $%
(|0\rangle +|1\rangle )/\protect\sqrt{2}$, $\protect\rho _{B}$, and $|%
\protect\varphi_{A}\rangle$, respectively. The index register is a single
qubit used as control qubit and to perform mQFT; the target register is used
to represent the state of subsystem \textit{B}; and the interacting register
is used to represent the state of subsystem \textit{A,} which interacts with 
\textit{B}. The circuit in the dotted square in (a) is used for obtaining
the $n$th binary digit of the real part of the phase factor. In part~(b), a
sequence of circuits inside the dashed squares is used to replace the
circuit in the dashed square in part~($a$), in order to resolve from the ($%
n-1$)th to the first binary digits of the real part of the phase factor.
Here, $H$ is the Hadamard gate, $Q_{k}$ is a single-qubit rotation as
defined in Eq.~($27$) and $R_{k}$ is a single-qubit rotation in quantum
Fourier transform. Part~(c) shows the circuit for the \textquotedblleft $k$%
th\textquotedblright\ partial measurement $W(k)$ with period $\protect\tau $%
. In the circuit, for the unitary transformation $U=\exp(-iHt)$, we set $t$
such that all projective measurements are performed successfully on
subsystem \textit{A} while the unitary transformation of the whole system
evolves for time $t$.}
\end{figure}

\section{Examples of measurement-based phase estimation}

\label{example}

\subsection{Phase estimation for the Jaynes-Cummings Hamiltonian}

Now we use a simple model to show how MPEA works. We consider here a quantum
system consisting of two subsystems \textit{A} and \textit{B}, where \textit{%
B} contains two non-interacting spin qubits, \textit{B}$_{\text{1}}$ and 
\textit{B}$_{\text{2}}$; and subsystem \textit{A }is a photon. The whole
system is described by the Jaynes-Cummings Hamiltonian~\cite{wu} 
\begin{equation}
H=w_{0}b^{\dag }b+w_{1}(\sigma _{1}^{z}+\sigma _{2}^{z})+\frac{J}{2}%
[b(\sigma _{1}^{+}+\sigma _{2}^{+})+b^{\dagger }(\sigma _{1}^{-}+\sigma
_{2}^{-})],
\end{equation}%
where $b$~($b^{\dagger }$) is a bosonic annihilation~(creation) operator of
the photons. Consider the case $w_{0}=w_{1}$, and perform projective
measurements in the basis of a single photon state $|\varphi _{A}\rangle
=|1\rangle $. Then we have 
\begin{eqnarray}
V_{B}(\tau ) &=&\text{diag}\biggl\{1,\bigl[3+2\cos (\sqrt{10}\tau J)\bigr]%
/5\times e^{-2iw_{0}\tau },  \nonumber \\
&&\cos (\sqrt{6}\tau J)e^{-iw_{0}\tau },\cos (\sqrt{2}\tau J)\biggr\},
\end{eqnarray}%
in the ordered basis $\bigl\{|1,s\rangle ,|1,t_{+}\rangle ,|1,t_{0}\rangle
,|1,t_{-}\rangle \bigr\}$, where 
\begin{eqnarray}
|s\rangle &=&\frac{1}{\sqrt{2}}(|01\rangle -|10\rangle ),\text{ \ \ }%
|t_{+}\rangle =|11\rangle ,  \nonumber \\
|t_{0}\rangle &=&\frac{1}{\sqrt{2}}(|01\rangle +|10\rangle ),\text{ \ \ }%
|t_{-}\rangle =|00\rangle .
\end{eqnarray}%
Let $\tau =1/2$ and $w_{0}=w_{1}=J=1$, then we construct an evolution matrix 
$V_{B}(\tau )$ as 
\begin{eqnarray}
V_{B}(\tau ) &=&\text{diag}\biggl\{1,\bigl[3+2\cos (\sqrt{10}/2)\bigr]%
/5\times e^{-i},  \nonumber \\
&&\cos (\sqrt{6}/2)e^{-i/2},\cos (\sqrt{2}/2)\biggr\}.
\end{eqnarray}

On the MPEA circuit, now let us prepare the target register in a mixed
state: 
\begin{equation}
\rho _{B}=\frac{1}{4}\biggl(|s\rangle \langle s|+|t_{+}\rangle \langle
t_{+}|+|t_{0}\rangle \langle t_{0}|+|t_{-}\rangle \langle t_{-}|\biggr),
\end{equation}%
and the interacting register in state $|\varphi _{A}\rangle =|1\rangle $.
Then implement the controlled Hamiltonian of Eq.~($28$) for a time $t$\
during which the projective measurements on the interacting register are
performed successfully. After a series of projective measurements, in the
basis of $|\varphi _{A}\rangle =|1\rangle $, on the interacting register,
the state of the target register evolves to a singlet state $|s\rangle $,
which corresponds to the largest eigenvalue of $V_{B}(\tau )$. We resolve
the corresponding phase as zero, thus its eigenvalue is one. The survival
probability, $P^{\left( \tau \right) }\left( m\right) $, of the state $%
|\varphi _{A}\rangle =|1\rangle $ on the interacting register after $m$
successful measurements, and the fidelity, $F^{\left( \tau \right) }(m)$,
for the target register to be in state $|u_{\max }\rangle $, are shown in
Fig.~$3$. The fidelity $F^{\left( \tau \right) }(m)$\ is defined as%
\begin{equation}
F^{\left( \tau \right) }\left( m\right) =\frac{\langle u_{\max }|\rho
_{B}^{(\tau )}(m)|u_{\max }\rangle }{\langle u_{\max }|u_{\max }\rangle }
\end{equation}%
Since $F^{\left( \tau \right) }\left( m\right) $ is close to one, the
success probability is determined by $P^{\left( \tau \right) }\left(
m\right) $. 
\begin{figure}[tbp]
\includegraphics[width=0.9\columnwidth, clip]{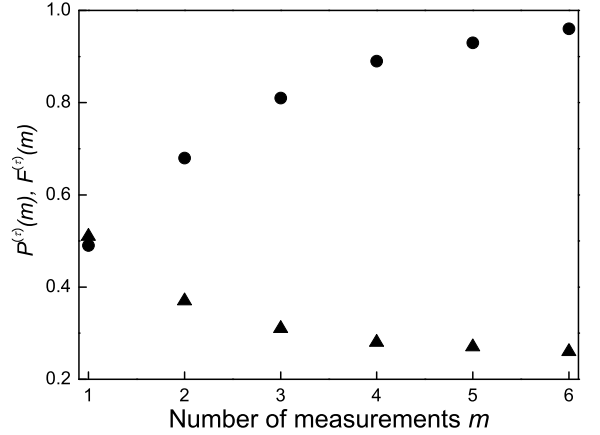} 
\caption{Survival probability $P(m)$~($\blacktriangle $) and fidelity $F(m)$%
~($\bullet $) for $|\protect\varphi _{A}\rangle =|1\rangle $ versus $m$, the
number of successful measurement, for the Jaynes-Cummings model.}
\end{figure}

If we prepare the target register in a pure initial state that is close to
an eigenstate of the matrix $V_{B}$, by applying MPEA, the state of the
target register can evolve to other eigenstates of $V_{B}$. Then we can also
obtain the corresponding eigenvalues of $V_{B}$.

For example, applying MPEA to the above system and preparing the target
register in state $|t_{+}\rangle $, by performing projective measurements
with $|\varphi _{A}\rangle =|1\rangle $ on the interacting register, the
state of the target register would remain in the $|t_{+}\rangle $ state. We
can retrieve the real part of the phase factor of the corresponding
eigenvalue up to an accuracy of $2$, $8$ and $16$ binary digits,
respectively, and obtain the eigenvalues of the matrices $V_{B}$ as $\exp
[-0.5177-i2\pi (0.25\pm 0.25)]$, $\exp [-0.5177-i2\pi (0.160\pm 0.008)]$,
and $\exp [-0.5177-i2\pi (0.15918\pm 0.00003)]$, assuming we have already
obtained the imaginary part of the eigenvalue of $V_{B}$\ through projective
measurements. The true eigenvalue is $\exp (-0.5177-i)$, which is quite
close.

To implement a controlled unitary evolution on the MPEA circuit, we set the
control qubit as a single spin and label it as subsystem \textit{C}. Thus,
the controlled Hamiltonian of the whole system becomes 
\begin{eqnarray}
\widetilde{H} &=&\frac{1}{2}(1-\sigma _{C}^{Z})H=\frac{1}{2}(1-\sigma
_{C}^{Z})\biggl\{w_{0}b^{\dag }b+w_{1}(\sigma _{1}^{z}+\sigma _{2}^{z})+ 
\nonumber \\
&&\frac{J}{2}\big[b(\sigma _{1}^{+}+\sigma _{2}^{+})+b^{\dagger }(\sigma
_{1}^{-}+\sigma _{2}^{-})\big]\biggr\}.
\end{eqnarray}%
This Hamiltonian contains three-body interactions and cannot be implemented
directly. One could decompose the three-body interaction into two-body
interactions~\cite{som, tseng} and then implement the two-body interaction.
In general, an arbitrary unitary matrix $U=\exp \left( -iHt\right) $ can be
decomposed \cite{mot, kha} into tensor products of unitary matrices of $%
4\times 4$\ and $2\times 2$, which correspond to two- and single-qubit
operations respectively, and can be implemented on a universal quantum
computer.

\subsection{Phase estimation for the axial symmetry model}

For another example, we consider the axial symmetry model~\cite{wu}. This is
relevant for quantum information processing in solid state~\cite{moz, ima,
qui} and atomic~\cite{zheng} systems. The quantum system is composed of two
subsystems \textit{A} and \textit{B}, where \textit{B} contains two
non-interacting spins, and subsystem \textit{A }contains\textit{\ }a single
spin interacting with subsystem \textit{B}. The Hamiltonian for the whole
system is~\cite{wu} 
\begin{equation}
H=\frac{J}{2}\bigl[X(X_{1}+X_{2})+Y(Y_{1}+Y_{2})\bigr],
\end{equation}%
where $X$ and $Y$ are the Pauli operators. By performing projective
measurements on subsystem \textit{A} in the basis of the $\sigma _{z}$%
-eigenvector, then in the basis $\{|s\rangle ,|t_{+}\rangle ,|t_{0}\rangle
,|t_{-}\rangle \}$, we obtain 
\begin{equation}
V_{B}(\tau )=\text{diag}\biggl\{1,1,\cos \left( \sqrt{2}\tau J\right) ,\cos
\left( \sqrt{2}\tau J\right) \biggr\},
\end{equation}%
operating on subsystem \textit{B}. If we prepare the initial state of the
target register in state $|t_{0}\rangle $, then the fidelity of the target
register to be in state $|t_{0}\rangle $ is $1$ after performing a number of
successful measurements on the interacting register. For the case $J=2$ and $%
\tau =1$, the corresponding eigenvalue is $-0.951363$. The success
probability of the successful measurement on the interacting register versus
the number of measurements on the interacting register is shown in Fig.~$4$.
From that figure, we can see that even for $10$ successful measurements, we
can still have a success probability of $0.37$. 
\begin{figure}[tbp]
\includegraphics[width=0.9\columnwidth, clip]{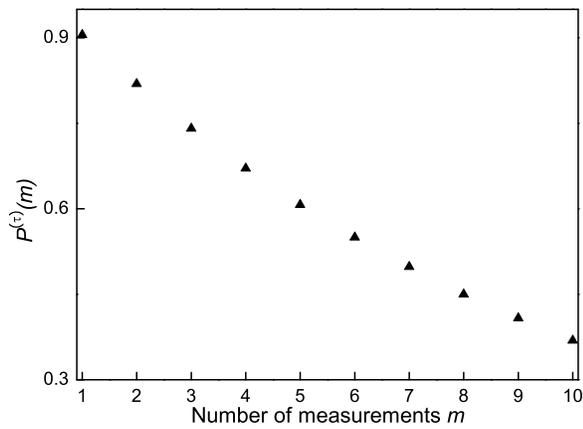} 
\caption{Survival probability $P(m)$ for $|\protect\varphi _{A}\rangle
=|1\rangle $ versus $m$, the number of successful measurements, by using the 
$\protect\sigma _{z}$-eigenstate as the measurement basis for the axial
symmetry model.}
\end{figure}

\section{Discussion}

\label{diss}

On a quantum computer, a unitary matrix can be efficiently represented,
i.e., for a unitary matrix of dimension $2^{s}$, only $s$ qubits are needed
to represent it on a quantum computer. In this paper, we have tried to
represent a non-unitary matrix on a quantum system by performing periodic
projective measurements. Whether an arbitrary matrix can be constructed
using this technique still remains an open problem, and this would be a
subject for future study.

\subsection{Implementation of the controlled nonunitary transformation}

In the conventional PEA, the phase factor is resolved through a quantum
Fourier transform. To resolve the binary expansion of the phase, up to $n$
binary digits, one has to implement $n$ controlled unitary transformations
in successive binary powers: $C$-$U^{2^{n-1}}$, $C$-$U^{2^{n-2}}, \cdots, C$-%
$U^{2^{0}}$.

In the MPEA approach of using mQFT combined with projective measurements to
obtain the eigenvalues of $V_{B}$, we need to implement the controlled
transformations in successive binary powers: $C$-$V_{B}^{2^{n-1}}\!\left(%
\tau \right) $, $C$-$V_{B}^{2^{n-2}}\!\left( \tau \right), \cdots, C$-$%
V_{B}^{2^{0}}\!\left(\tau \right) $, and followed by the corresponding mQFT
circuit as shown in Fig.~$2$. The controlled transformations $C$-$%
V_{B}^{2^{n-1}}\!\left( \tau \right) $, $C$-$V_{B}^{2^{n-2}}\!\left(\tau
\right)$, $\cdots $, $C$-$V_{B}^{2^{0}}\!\left(\tau \right)$, are achieved
by implementing the controlled Hamiltonian on the whole system only once and
during a time $t$ until the successful measurements on the interacting
register finish, and performing measurements $W(n-1), W(n-2), \cdots, W(0)$,
i.e., a series of periodic measurements (each one separated by the time
interval $\tau $) for $2^{n-1},$ $2^{n-2}, \cdots, 2^{0}$ times on the
interacting register, respectively.

\subsection{Success probability}

The success probability of the MPEA is $F^{\left( \tau \right) }\!\left(
m\right) P^{\left( \tau \right) }\!\left( m\right) $, where $F^{\left( \tau
\right) }\!\left( m\right) $ is the fidelity of the state on the target
register to be in the eigenstate of $V_{B}\!\left(\tau \right) $ after
performing $m$ successful measurements, and $P^{\left( \tau \right)
}\!\left( m\right) $ is the probability of performing $m$ successful
measurements on the interacting register. Note that $P^{\left( \tau \right)
}\!\left( m\right) $ depends on $|\lambda _{\max }|$, and also on the
initial guess of the state on the target register as shown in Eq.~($13$).
Since $F^{\left( \tau \right) }\!\left( m\right) $ is close to one as the
number of successful measurements $m$ increases, the success probability is
determined by $P^{\left( \tau \right) }\!\left( m\right) $.

It must be emphasized that the present quantum algorithm is designed for
systems with large $D_{B}$~(the dimension of subsystem $B$), when the
corresponding classical algorithms for obtaining the eigenvalues of $%
V_{B}\!\left( \tau \right) $ become so expensive that it is impossible to
implement on a classical computer. The efficiency of our algorithm \emph{does%
} depend on $P^{\left( \tau \right) }\!\left( m\right) $. Note that the
success probability of projective measurements on the interacting register
decreases exponentially in terms of $m$ when $|\lambda _{\max }|<1$. This is
not an essential obstacle because this exponential decrease can be overcome
by running the algorithm for a number of times to prepare a large but \emph{%
fixed} number of copies of the index qubit state as shown in Eq.~($20$). In
the QST approach to obtain $\lambda _{\max }$, the measurement errors of QST
obey the central limit theorem. Accurate results can be obtained by
preparing a larger ensemble of the single qubit states. The tomographic
estimation converges with statistical error that decreases as $N^{-1/2}$,
where $N$ is the number of copies prepared in the QST and is not relevant to 
$D_{B}$.

Also, in the approach of using single-qubit QST to obtain the eigenvalues of 
$V_{B}\!(\tau )$, we prepare a number of copies of the index qubit state as
shown in Eq.~$(20)$. If we have a good initial guess of the eigenstate of $%
V_{B}\!(\tau )$, then, as shown in the second example, we can still obtain a
high success probability ($F^{\left( \tau \right) }\!\left( m\right)
P^{\left( \tau \right) }\!\left( m\right) $) for the algorithm and this does
not require a large $m$.

The other eigenvalues of $V_{B}$ can be obtained by setting the initial
state of the target register in a pure state. If the overlap of the initial
guess of the eigenstate with the real eigenstate is not exponentially small
and $m$ is a fixed number, the success probability, $F^{\left( \tau \right)
}\!\left( m\right) P^{\left( \tau \right) }\!\left( m\right) $, for
preparing a index qubit state as shown in Eq.~($20$) is not exponentially
small. Then each copy of the index qubit state can be prepared in a
polynomial number of trials.

\subsection{Efficiency for projective measurements}

Another issue that needs to be addressed is the efficiency for implementing
the projective measurement $M=|\varphi ^{A}\rangle \langle \varphi ^{A}|$,
which is linked to the efficiency of constructing the non-unitary matrix $%
V_{B}\left( \tau \right) $, therefore connected to the efficiency of the
algorithm. Since the measurement $M$ is a non-unitary process, it cannot be
implemented deterministically. Also, a number of projective measurements are
required in MPEA, and thus the overall efficiency of the algorithm might be
affected. To deal with this problem, we can design a scheme such that
subsystem $A$ can have a simple structure, containing either a single qubit
or a few qubits, by controlling the interaction between the subsystems. Then
the implementation of the measurement on subsystem $A$ will be simple. The
measurement performed on $A$ does not depend on the qubit number $n_{B}$ of
the subsystem $B$, on which the matrix $V_{B}$ is constructed. Therefore,
the measurement on $A$ can avoid the exponential scaling with respect to the
size of subsystem $B$. Note that the corresponding classical algorithms
scale as $2^{n_{B}}$.

\section{Conclusion}

\label{con}

We have presented a measurement-based quantum phase estimation algorithm to
obtain the eigenvalues and the corresponding eigenvectors of non-unitary
matrices. In MPEA, we implement the unitary transformation of the whole
system only once; the non-unitary matrix is constructed as the evolution
matrix on the target register. By performing periodic projective
measurements on the interacting register, the state of the target register
is driven automatically to a pure state of the transformation matrix. Using
single-qubit state tomography and mQFT combined with single-qubit projective
measurements, we can obtain the complex eigenvalues of the non-unitary
matrix. The success probability of the algorithm and the efficiency of
constructing the matrix $V_{B}\left( \tau \right) $ have been discussed.
This algorithm can be used to study open quantum system and in developing
other new quantum algorithms.

\begin{acknowledgements}
FN acknowledges partial support from DARPA, Air Force Office for
Scientific Research, the Laboratory of Physical Sciences, National
Security Agency, Army Research Office, National Science Foundation
grant No.~0726909, JSPS-RFBR contract No.~09-02-92114,
Grant-in-Aid for Scientific Research~(S), MEXT Kakenhi on Quantum
Cybernetics, and Funding Program for Innovative R\&D on
S\&T~(FIRST). LAW thanks the support of the Ikerbasque Foundation.
\end{acknowledgements}

\end{document}